\documentclass[prb,twocolumn,showpacs,amsmath,amssymb]{revtex4}

\usepackage{graphicx}
\usepackage{dcolumn}
\usepackage{bm}
\usepackage{amsmath}

\begin{document}
\title{%
Magnetic-Field Dependences of
Thermodynamic Quantities \\
in the Vortex State of Type-II Superconductors}

\author{Koichi Watanabe}
\affiliation{Division of Physics, Hokkaido University, Sapporo 060-0810, Japan}

\author{Takafumi Kita}
\affiliation{Division of Physics, Hokkaido University, Sapporo 060-0810, Japan}

\author{Masao Arai}
\affiliation{
National Institute for Materials Science, Namiki 1-1, Tsukuba, Ibaraki 305-0044, Japan}

\date{\today}

\begin{abstract}
We develop an alternative method to solve the Eilenberger equations numerically
for the vortex-lattice states of type-II superconductors.
Using it, we clarify the magnetic-field and impurity-concentration dependences of 
the magnetization,
the entropy, the Pauli paramagnetism, and the mixing of higher Landau levels
in the pair potential for two-dimensional $s$- and $d_{x^2-y^2}$-wave superconductors
with the cylindrical Fermi surface.
\end{abstract}
\maketitle

\section{Introduction}

Recent experiments\cite{Sologubenko01,Boaknin03,Izawa03,
Aoki04,Deguchi04} have shown that magnetic-field dependences 
of thermodynamic quantities in the vortex state of type-II superconductors
provide unique information on the pairing symmetry and
gap anisotropy.
On the theoretical side, however, calculations of those quantities 
still remain a fairly difficult task to perform.
The quasiclassical equations derived by Eilenberger\cite{Eilenberger}
provide a convenient starting point for this purpose.
Pesch\cite{Pesch75} obtained a compact analytic solution to them
based on the lowest Landau-level approximation for the pair potential.
It has been used recently to discuss 
the field dependences of the thermal conductivity,\cite{Vekhter99}
the density of states,\cite{Dahm02,Graser04,Udagawa04} and
thermodynamic quantities.\cite{Kusunose04}
On the other hand, Klein\cite{Klein87,Klein89} 
obtained a full numerical solution for the vortex-lattice state
using a standard procedure to solve ordinary
differential equations.
This numerical approach has been used extensively by Ichioka {\em et al}.\
\cite{Ichioka96,Ichioka97,Ichioka99a,Ichioka99b,Ichioka02,Ichioka03,
Ichioka04a,Ichioka04b,Ichioka04c} 
to clarify the field dependences of the pair potential and the density of states
for type-II superconductors with various energy-gap structures.
It should be noted, however, that those numerical studies 
all adopted simplified model Fermi surfaces
instead of complicated Fermi surfaces for real materials.
Indeed, recent theoretical studies\cite{Graser04,RS90,Kita04,Arai04} 
have clarified that detailed Fermi-surface structures are indispensable
for the quantitative description of the vortex state.

With these backgrounds, 
we here develop an alternative method to solve the 
Eilenberger equations for the vortex-lattice states.
A key point lies in expanding the pair potential and the quasiclassical $f$ function
in the basis functions of the vortex-lattice states,
thereby transforming the differential equations into algebraic equations.
This method has been powerful for (i) solving the Ginzburg-Landau
equations\cite{Kita98} and 
the Bogoliubov-de Gennes equations\cite{NMA95,Kita98-2,YK99,YK02}
and (ii) obtaining quantitative agreements
on the upper critical field $H_{c2}$ of Nb, NbSe$_{2}$, and MgB$_{2}$
with Fermi surfaces from 
first-principles electronic-structure calculations.\cite{Kita04,Arai04}
Thus, the method may be more advantageous for the calculations of
thermodynamic quantities in finite magnetic fields 
when realistic Fermi surfaces are used as inputs.
It will also be convenient for microscopically calculating 
responses of the vortex-lattice state
to external perturbations,\cite{Kita04b}
such as vortex-lattice oscillations.

This method is applied here to calculate magnetic-field dependences
of the magnetization, the entropy, the Pauli paramagnetism, and the pair potential
at various temperatures for the two-dimensional $s$- and $d_{x^2-y^2}$-wave 
superconductors in the clean and dirty limits.
These quantities have been obtained near $H_{c2}$ 
for the $s$-wave pairing.\cite{Kita03,Kita04c}
Our purpose here is to clarify the overall field dependence
of those quantities.

This paper is organized as follows.
Section II gives the formulation. 
Section III presents numerical results.
Section IV summarizes the paper.
We use $ k_{\rm B}\!=\! 1$ throughout. 

\section{Formulation}

\subsection{Eilenberger equations}

We take the external magnetic 
field ${\bf H}$ along the $z$ axis 
and express the vector potential as
\cite{Eilenberger,Lasher,Marcus,Brandt,Doria,Kita98}
\begin{equation}
{\bf A}({\bf r}) = Bx\hat{{\bf y}} + \tilde{{\bf A}}({\bf r}) \: .
\label{eq:vector potential}
\end{equation} 
Here $B$ is the average flux density 
produced jointly by the external current and the internal supercurrent,
and $\tilde{{\bf A}}$ is the spatially varying 
part of the magnetic field 
satisfying $\int {\bm \nabla}\!\times\! \tilde{{\bf A}} \, d{\bf r}\!=\!{\bf 0}$.
We choose the gauge such that
$
{\bm \nabla}\!\cdot \! \tilde{{\bf A}} \!=\! {\bf 0}$.

The Eilenberger equation for the even-parity pairing
without Pauli paramagnetism
is given by\cite{Eilenberger,Kita04}
\begin{subequations}
\label{Eilens}
\begin{equation}
\left(\!\varepsilon_{n}+\!\frac{\hbar}{2\tau}\langle g\rangle
\!+\!\frac{1}{2}\hbar{\bf v}_{{\rm F}}\!\cdot\!{\bm \partial}\!\right)\! f=
\left(\!\Delta \phi \!+\!\frac{\hbar}{2\tau}\langle f\rangle\!\right)\! g \, .
\label{Eilen}
\end{equation}
Here $\varepsilon_{n}\!=\!(2n+ 1)\pi T$ $(n\!=\!0,\pm 1,\pm 2,\cdots)$
is the Matsubara energy with $T$ the temperature, 
$\tau$ is the relaxation time by nonmagnetic impurity scattering,
and $\langle\cdots\rangle$ denotes the Fermi-surface average:
\begin{eqnarray*}
\langle g \rangle \equiv
\int\!\! dS_{{\rm F}} \frac{
g (\varepsilon_{n},{\bf k}_{\rm F},{\bf r})}{(2\pi)^{3}N(0)|{\bf v}_{{\rm F}}|} \, ,
\end{eqnarray*}
with $dS_{{\rm F}}$ an infinitesimal area on the Fermi surface,
$N(0)$ the density of states per spin and per unit volume
at the Fermi energy in the normal state, and ${\bf v}_{\rm F}$ the Fermi velocity.
The operator ${\bm \partial}$ in Eq.\ (\ref{Eilen}) is defined by
\begin{eqnarray*}
{\bm \partial}\equiv {\bm \nabla}-i\frac{2\pi}{\Phi_{0}} {\bf A} \, ,
\end{eqnarray*}
with $\Phi_{0}\!\equiv\! hc/2e$ the flux quantum,\cite{comment1}
$\Delta({\bf r})$ is the pair potential,
and $\phi({\bf k}_{\rm F})$ specifies the gap anisotropy satisfying
$\langle\phi({\bf k}_{\rm F})\rangle\!=\! 1$.
Finally, the quasiclassical Green's functions $f$ and $g$
for $\varepsilon_{n}\!>\! 0$ are connected by
$g\!=\!(1\!-\! ff^{\dagger})^{1/2}$
with $f^{\dagger}(\varepsilon_{n},{\bf k}_{\rm F},{\bf r})\!=
\!f^{*}(\varepsilon_{n},-{\bf k}_{\rm F},{\bf r})$.

Equation (\ref{Eilen}) has to be solved simultaneously
with the self-consistency equation for the pair potential and
the Maxwell equation for $\tilde{\bf A}$, which are 
given respectively by
\begin{equation}
\Delta({\bf r}) \ln \!\frac{T_{c0}}{T}= 2\pi T \! \sum_{n=0}^{\infty}
\left[\frac{\Delta({\bf r})}{\varepsilon_{n}}-\langle 
\phi({\bf k}_{\rm F})f(\varepsilon_{n},{\bf k}_{\rm F},{\bf r})\rangle 
\right]  ,
\label{pair}
\end{equation}
\begin{equation}
-\nabla^{2} \tilde{{\bf A}}({\bf r}) = -i\frac{16\pi^{2}e N(0)T}{c} 
\sum_{n=0}^{\infty} \langle {\bf v}_{{\rm F}} 
g (\varepsilon_{n}, {\bf k}_{{\rm F}}, {\bf r}) \rangle \:, 
\label{eq:Maxwell}
\end{equation}
\end{subequations}
with $T_{c0}$ the transition temperature
for $\tau\!=\!\infty$.

Finally, the free-energy functional corresponding to
Eq.\ (\ref{Eilens}) is given by\cite{Eilenberger,Kita03}
\begin{eqnarray}
&&\hspace{-7mm}F_{\rm s}
=F_{\rm n}+\int\!d{\bf r}
\biggl\{ \frac{
({\bm \nabla}\!\times\!{\bf A})^{2}}{8\pi}+N(0)|\Delta({\bf r})|^{2}\ln\frac{T}{T_{c0}}
\nonumber \\
&&+2\pi T N(0) \sum_{n=0}^{\infty}\left[\frac{|\Delta({\bf r})|^{2}}{\varepsilon_{n}}-
\langle I(\varepsilon_{n},{\bf k}_{\rm F},{\bf r})\rangle \right]
\biggr\} ,
\label{F}
\end{eqnarray}
where $F_{n}$ is the free energy in the normal state 
and $I$ is defined by
\begin{eqnarray}
&&\hspace{-7mm}I\equiv \Delta^{\! *}f\!+\!\Delta f^{\dagger} 
+2\varepsilon_{n}(g\!-\!1)
+\hbar\,\frac{f\langle f^{\dagger}\rangle\!+\!\langle f\rangle f^{\dagger}}{4\tau}
\nonumber \\
&&\hspace{-1mm}
+\hbar\,\frac{g\langle g\rangle\!-\! 1}{2\tau}
-\hbar\,\frac{f^{\dagger}\,{\bf v}_{\rm F}\!\cdot\!
{\bm \partial}f
-f\,{\bf v}_{\rm F}\!\cdot\!
{\bm \partial}^{*}f^{\dagger}}
{2(g\!+\!1)} \, .
\label{I}
\end{eqnarray}
Indeed, functional differentiations of Eq.\ (\ref{F})
with respect to $f$, $\Delta$, and $\tilde{\bf A}$ lead to Eqs.\
(\ref{Eilen}), (\ref{pair}), and (\ref{eq:Maxwell}), respectively.

\subsection{Operators and basis functions}

We first express the gradient operator in Eq.\ (\ref{Eilen}) as
\begin{equation}
{\bf v}_{{\rm F}}\!\cdot\!{\bm \partial}
= \frac{{\bar{v}_{{\rm F}+}^{*} (a\!+\!\tilde{A}) 
- \bar{v}_{{\rm F}+} (a^{\dagger}\!+\!\tilde{A}^{*})}}{\sqrt{2}l_{\rm c}} \, .
\label{gradient}
\end{equation}
Here 
$a$ and $a^{\dagger}$ are the boson operators:
\begin{subequations}
\label{aAv}
\begin{equation}
\left[
	\begin{array}{c}
	a \\ a^{\dagger}
	\end{array}
\right]
=\frac{l_{{\rm c}}}{\sqrt{2}}
\left[
	\begin{array}{cc}
	c_{1} & ic_{2} \\
	-c_{1}^{*} & ic_{2}^{*}
	\end{array}
\right]
\left[
	\begin{array}{c}
	\nabla_{x} \\ \nabla_{y}-2\pi i Bx/\Phi_{0}
	\end{array}
\right] \, ,
\label{aa*}
\end{equation}
with $l_{\rm c}\!\equiv\! \sqrt{\Phi_{0}/2\pi B}$ and
$c_{1}c_{2}^{*}+c_{1}^{*}c_{2}\!=\! 2$, and
$\tilde{A}$ and $\bar{v}_{{\rm F}+}$ are defined by
\begin{equation}
\tilde{A}\equiv-i\frac{\sqrt{2}\pi l_{\rm c}}{\Phi_{0}}(c_{1}\tilde{A}_{x}
+ic_{2}\tilde{A}_{y}) \, ,
\label{tildeA}
\end{equation}
\begin{equation}
\label{v_+}
\bar{v}_{{\rm F}+}\equiv c_{2}v_{{\rm F}x}+i c_{1}v_{{\rm F}y} \, ,
\end{equation}
\end{subequations}
respectively.
The constants ($c_{1}$,$c_{2}$) can be fixed appropriately 
to make the subsequent calculations efficient.
A convenient choice\cite{Kita04}  is to impose the condition that 
the gradient term 
in the Ginzburg-Landau equation be expressed in terms of $a^{\dagger}a$
without using $aa$ and $a^{\dagger}a^{\dagger}$, i.e.,
the pair potential near $T_{c}$ be described 
in terms of the lowest Landau level only.
Alternatively, one may follow Graser {\em et al}.\ \cite{Graser04} 
to change them at every temperature and 
magnetic field so as to make the free-energy within the lowest-Landau-level
approximation smallest.

Using Eq.\ (\ref{aa*}), we can make up
a set of basis functions to describe arbitrary
vortex-lattice structures as\cite{Kita98}
\begin{eqnarray}
&&\hspace{-4mm} \psi_{N{\bf q}}({\bf r})=
\sqrt{\frac{2\pi l_{\rm c}}{c_{1}a_{2}\sqrt{\pi }\,V }}
\sum_{n=-{\cal N}_{{\rm f}}/2+1}^{{\cal N}_{{\rm f}}/2}
\exp\!\left[
i q_{y}\!\left(y+ \frac{l_{\rm c}^{2}q_{x}}{2}\right)\!\right]
\nonumber \\
&&\hspace{11mm}
\times\exp\!\left[
i \frac{na_{1x}}{l_{\rm c}^{2}}\!\left(y+
l_{\rm c}^{2}q_{x}- \frac{na_{1y}}{2}\right)
\!\right]
\nonumber \\
&&\hspace{11mm}
\times \exp\!\left[
-\frac{c_{1}c_{2}}{2}\left(\! 
\frac{x- l_{\rm c}^{2}q_{y}- na_{1x}}{c_{1}l_{\rm c}}\!\right)^{\!\! 2} \right] 
\nonumber \\
&&\hspace{11mm}\times 
\frac{1}{\sqrt{2^N N!}}H_{N}\!\!\left(\! 
\frac{x- l_{\rm c}^{2}q_{y}- na_{1x}}{c_{1}l_{\rm c}}\!\right) \, .
\label{basis}
\end{eqnarray}
Here $N\!=\!0,1,2,\cdots$ denotes the Landau level,
${\bf q}$ is an arbitrary chosen magnetic Bloch vector 
characterizing the broken translational symmetry of the vortex lattice
and specifying the core locations, and
$V$ is the volume of the system.
The quantities
$a_{1x}$, $a_{1y}$ and $a_{2}$ are the components of the basic vectors
${\bf a}_{1}$ and ${\bf a}_{2}$ in the $xy$ plane, respectively,
with ${\bf a}_{2}\!\parallel\!\hat{\bf y}$
and $a_{1x}a_{2}\!=\!2\pi l_{{\rm c}}^{2}$,
${\cal N}_{{\rm f}}^{2}$ denotes the number of the flux quantum 
in the system, and
$H_{N}(x)\!\equiv {\rm e}^{x^{2}}\!\left(-\frac{d}{dx}\!\right)^{\! N}
{\rm e}^{-x^{2}}$ is the Hermite polynomial.
The basis functions are both orthonormal and complete, satisfying
$a\psi_{N{\bf q}}\!=\!\sqrt{N}\psi_{N-1{\bf q}}$
and $a^{\dagger}\psi_{N{\bf q}}\!=\!\sqrt{N\!+\! 1}\psi_{N+1{\bf q}}$.

\subsection{Algebraic Eilenberger equations}

We now expand $\Delta$, $f$, and $\tilde{A}$ in the basis functions of the 
vortex lattice as
\begin{subequations}
\label{DfExpand}
\begin{equation}
\Delta({\bf r})= \sqrt{V}\sum_{N=0}^{\infty}\Delta_{N}\,\psi_{N{\bf q}}({\bf r}) \, ,
\label{DExpand}
\end{equation}
\begin{equation}
f(\varepsilon_{n},{\bf k}_{{\rm F}},{\bf r})
= \sqrt{V}\sum_{N=0}^{\infty}f_{N}(\varepsilon_{n},{\bf k}_{{\rm F}})\,
\psi_{N{\bf q}}({\bf r}) \, ,
\label{fExpand}
\end{equation}
\begin{equation}
\tilde{A}({\bf r})
= \sum_{{\bf K}\neq{\bf 0}}\tilde{A}_{\bf K}{\rm e}^{i{\bf K}\cdot{\bf r}} \, ,
\label{AExpand}
\end{equation}
\end{subequations}
where ${\bf K}$ is the reciprocal-lattice vector.\cite{Kita98}
Substituting Eq.\ (\ref{DfExpand}) into Eq.\ (\ref{Eilens}) and using the orthogonality
of the basis functions, Eq.\ (\ref{Eilens}) is transformed into a set
of algebraic equations for $\{f_{N}\}$, $\{\Delta_{N}\}$, and 
$\{\tilde{A}_{\bf K}\}$ as
\begin{subequations}
\label{Eilens2}
\begin{eqnarray}
&&\hspace{-10mm}
\varepsilon_{n}f_{N}+\bar{\beta}^{*} \sqrt{N\!+\! 1} f_{N+1}-
\bar{\beta} \sqrt{N}  f_{N-1}
\nonumber \\
&&\hspace{-10mm}=
\frac{1}{\sqrt{V}}\!\int \!\psi_{N{\bf q}}^{*}
\left(\!\Delta \phi g \!+\!\hbar\frac{\langle f\rangle g\!-\!\langle g\rangle f}{2\tau}
\!-\! \bar{\beta}^{*} \tilde{A} \!+\!
\bar{\beta} \tilde{A}^{*} \!\right)\! d{\bf r} \, ,
\label{Eilen2}
\end{eqnarray}
\begin{equation}
\Delta_{N} \ln \!\frac{T_{c0}}{T}= 2 \pi T \! \sum_{n=0}^{\infty}
\left[\frac{\Delta_{N}}{\varepsilon_{n}}-\langle 
\phi({\bf k}_{\rm F})f_{N}(\varepsilon_{n},{\bf k}_{\rm F})\rangle 
\right]  ,
\label{pair2}
\end{equation}
\begin{equation}
\tilde{A}_{\bf K} = -\frac{16\pi^{2}N(0)T}{(K l_{\rm c}B)^{2}V} 
\sum_{n=0}^{\infty} \int \langle \beta\, 
g (\varepsilon_{n}, {\bf k}_{{\rm F}}, {\bf r}) \rangle{\rm e}^{-i{\bf K}\cdot{\bf r}}
d{\bf r}
 \:, 
\label{eq:Maxwell2}
\end{equation}
\end{subequations}
with 
\begin{equation}
\bar{\beta}\equiv \frac{\hbar (c_{2}v_{{\rm F}x}\!+\!
i c_{1}v_{{\rm F}y})}{2\sqrt{2}l_{\rm c}} \, ,\hspace{3mm}
\beta\equiv \frac{\hbar (c_{1}v_{{\rm F}x}\!+\!
i c_{2}v_{{\rm F}y})}{2\sqrt{2}l_{\rm c}} \, .
\end{equation}
Together with the equation to determine $H_{c2}$
derived recently,\cite{Kita04}
the above coupled equations form a basis for efficient numerical
calculations of the Eilenberger equations for vortex-lattice states
with arbitrary Fermi-surface structures.

\subsection{Numerical procedures}

For a given vortex-lattice structure specified by the basic vectors
${\bf a}_{1}$ and ${\bf a}_{2}$ in Eq.\ (\ref{basis}),
the coupled equation (\ref{Eilens2}) may be solved iteratively 
in order of Eqs.\ (\ref{Eilen2}),
(\ref{pair2}), and (\ref{eq:Maxwell2}) 
by adopting a standard technique to solve nonlinear equations.\cite{NumRec}
A convenient starting point is to put
$\Delta_{N}^{(0)}\!=\!\delta_{N0}
\Delta(T)\sqrt{1\!-\! B/H_{c2}}$, $f_{N}\!=\! 0$, and
$\tilde{A}_{\bf K}\!=\! {\bf 0}$ on the right-hand 
of Eq.\ (\ref{Eilen2}), where $\Delta(T)$ is the angle-averaged
energy gap in zero field.
In this connection, it may be worth noting that the tridiagonal matrix 
constructed from the coefficients of $f_{N}$
on the left-hand side of Eq.\ (\ref{Eilen2}) can be 
inverted analytically.\cite{Kita04}
The Fermi-surface integrals $\langle \cdots\rangle$ can be performed 
as described in Sec.\ IV of Ref.\ \onlinecite{Kita04}.
In contrast, integrations over ${\bf r}$ in Eqs.\
(\ref{Eilen2}) and (\ref{eq:Maxwell2}) may be carried out 
as follows:
At the beginning of each calculation, 
we prepare $\psi_{N{\bf q}}({\bf r})$ and ${\rm e}^{i{\bf K}\cdot{\bf r}}$
at equally spaced $N_{\rm int}\!\times\!N_{\rm int}$
discrete points in a unit cell.
We then construct $\Delta({\bf r})$, $f(\varepsilon_{n},{\bf k}_{\rm F},{\bf r})$,
$\tilde{A}({\bf r})$, and $g\!=\!(1\!-\! ff^{\dagger})^{1/2}$
on those points by Eqs.\ (\ref{DfExpand}) with
restricting the summations to those satisfying
$N\!\leq\! N_{\rm c}$ and $|{\bf K}|\!\leq\! K_{\rm c}$,
where $N_{\rm c}$ and $K_{\rm c}$ are some cutoffs.
Now, the integrations can be performed by the trapezoidal rule; its
convergence is excellent for periodic functions.
Also, the summation over $n$ in Eq.\ (\ref{pair2}) is restricted in the
actual calculations to those satisfying
$|\varepsilon_{n}|\!\leq\!\varepsilon_{\rm c}$.
The convergence can be checked by increasing $N_{\rm int}$,
$N_{\rm c}$, $K_{\rm c}$, and $\varepsilon_{\rm c}$.
Finally, the vortex-lattice structure can be fixed
by requiring that the free energy (\ref{F}) be minimum.

\subsection{Thermodynamic quantities}

Once $\Delta$, $f$, and $\tilde{\bf A}$ are determined as above,
we can calculate thermodynamic quantities of the vortex-lattice state.
Specifically, the magnetization $M$ due to supercurrent and
the entropy $S_{\rm s}$
are obtained by\cite{Kita04c}
\begin{subequations}
\label{thermo}
\begin{eqnarray}
&&\hspace{-12mm}
-4\pi M =  \frac{1}{BV}\! \int \! d{\bf r} \biggl[\,(\nabla \! \times\! \tilde{{\bf A}})^{2}
\nonumber \\
&& \hspace{-4mm}+ 2\pi^{2}N(0) T  \sum_{n=0}^{\infty}  \biggl\langle 
\frac{f^{\dagger} {\bf v}_{{\rm F}} \!\cdot\! {\bm \partial} f 
- f {\bf v}_{{\rm F}} \!\cdot\! {\bm \partial}^{\ast} f^{\dagger}}
{g \!+\! 1} 
\biggr\rangle \, ,
\label{eq:magnetization}
\end{eqnarray}
\begin{eqnarray}
&&\hspace{-10mm} 
S_{\rm s}=S_{\rm n} -\frac{N(0)}{T}\!
\int \! d{\bf r} \biggl\{ |\Delta({\bf r})|^{2}
\nonumber \\
&& \hspace{-5mm} 
-2\pi T \sum_{n=0}^{\infty} \left[
\langle I(\varepsilon_{n}, {\bf k_{{\rm F}}}, {\bf r}) \rangle
+2\varepsilon_{n}\langle g\!-\!1 \rangle \right] \biggr\} \:,
 \label{eq:entropy}
\end{eqnarray}
respectively, where $S_{\rm n}\!\equiv\! 2\pi^{2}VN(0)T/3$ 
is the entropy in the normal state and $I$
is defined by Eq.\ (\ref{I}).
Also, when it is much smaller than the diamagnetism by supercurrent,
the magnetization $M_{{\rm sP}}$ due to Pauli paramagnetism 
can be calculated by\cite{Kita04c}
\begin{equation}
M_{{\rm sP}} = M_{{\rm nP}}\left[1
-\frac{2\pi T}{V}\sum_{n=0}^{\infty} \int \! d{\bf r} \,
\frac{\partial \langle g \rangle}{\partial \varepsilon_{n}}
\biggl(\!1+ \frac{\nabla \!\times\! \tilde{\bf A}}{B}\! \biggr)^{\!\! 2}\, \right] \:,
\label{eq:spin susceptibility}
\end{equation}
\end{subequations}
where $M_{{\rm nP}}\!\equiv\! 2VN(0)\mu_{\rm B}^{2} B$ with $\mu_{\rm B}$ the
Bohr magneton.
The quantity
\begin{eqnarray}
\frac{\partial g}{\partial \varepsilon_{n}}=
-\frac{1}{2(1\!-\! ff^{\dagger})^{1/2}}
\left(\! f\frac{\partial f^{\dagger}}{\partial \varepsilon_{n}}\!
+\!\frac{\partial f}{\partial \varepsilon_{n}}f^{\dagger}\!\right)
\end{eqnarray}
in Eq.\ (\ref{eq:spin susceptibility}) may be obtained
either by numerical differentiations or directly from the equation
of differentiating Eq.\ (\ref{Eilen}) with respect to $\varepsilon_{n}$:
\begin{eqnarray}
&&
\left(\!\varepsilon_{n} + \frac{\hbar}{2\tau} \langle g \rangle
+\frac{1}{2} \hbar{\bf v}_{{\rm F}} \cdot {\bm \partial}\!\right)
\frac{\partial f}{\partial \varepsilon_{n}}
+\left(\!1+\frac{\hbar}{2\tau}\frac{\partial \langle g \rangle}{\partial \varepsilon_{n}}
\!\right)f
\nonumber \\ 
&&=\left(\!\Delta \phi+ \frac{\hbar}{2\tau}\langle f \rangle\!\right)
\frac{\partial g}{\partial \varepsilon_{n}}
+\frac{\hbar}{2\tau}\frac{\partial \langle f \rangle}{\partial \varepsilon_{n}}g
\:. \label{eq:Eilenberger different}
\end{eqnarray}
This equation can be solved similarly as Eq.\ (\ref{Eilen}).

\section{Results}

We now present numerical results for two-dimensional systems
with the cylindrical Fermi surface
which is placed in the $xy$ plane perpendicular to ${\bf H}$.
We have considered a couple of energy gaps:
\begin{equation}
\phi({\bf k}_{\rm F})=
\left\{
\begin{array}{cl}
\vspace{1mm}
1 & \hspace{3mm} :\mbox{$s$-wave}
\\
\sqrt{2}\,(\hat{k}_{{\rm F}x}^{2}\!-\!\hat{k}_{{\rm F}y}^{2}) 
& \hspace{3mm}:\mbox{$d_{x^{2}-y^{2}}$-wave}
\end{array}
\right. .
\end{equation}
Then it is convenient to set $c_{1}\!=\!c_{2}\!=\! 1$ in Eq.\ (\ref{aAv}).
There is another parameter in the system corresponding to the Ginzburg-Landau 
parameter $\kappa$. 
We have fixed it by
\begin{equation}
\kappa_{0}\equiv
\frac{(\hbar c/2e)\Delta(0)}
{\sqrt{4\pi N(0)}\,\hbar^{2}{v}_{{\rm F}}^{2}} 
=\left\{\begin{array}{cl}
\vspace{1mm}
10 & \hspace{3mm}\mbox{: $s$-wave}
\\
7 & \hspace{3mm}\mbox{: $d_{x^2-y^2}$-wave}
\end{array}
\right.
\, ,
\label{kappa_0}
\end{equation}
with $\Delta(0)$ the angle-averaged energy gap at $T\!=\!0$.
It follows from Eq.\ (46) of Ref.\ \onlinecite{Kita03}
that Eq.\ (\ref{kappa_0}) corresponds to $\kappa\!=\! 49$ and $40$
for the $s$- and $d_{x^2-y^2}$-wave pairings in the clean limit,
respectively,
and $\kappa\!=\! 1300$ for the $s$-wave pairing 
with $\tau\!=\!0.01\hbar/\Delta(0)$.
The vortex-lattice structure has been fixed as hexagonal (square) 
for the $s$-wave ($d_{x^{2}-y^{2}}$-wave) pairing so that
finite contributions in the expansion (\ref{DExpand}) 
come only from $N\!=\!0,6,12,\cdots$ ($N\!=\!0,4,8,\cdots$) Landau levels.

\begin{figure}[t]
\begin{center}
\includegraphics[width=0.75\linewidth]{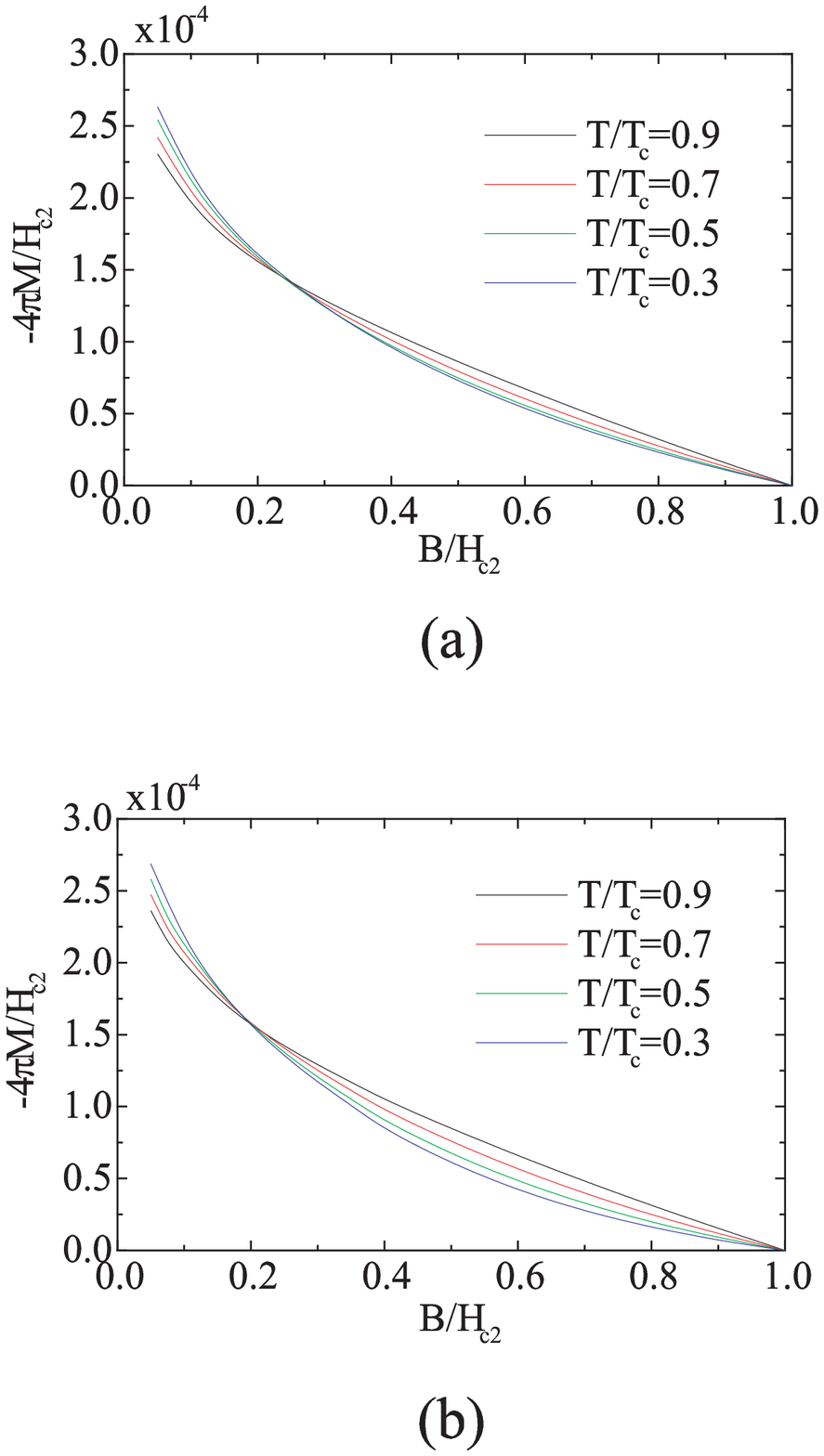}
\caption{The magnetization by supercurrent as a function of 
$B/H_{{\rm c2}}$ in the clean limit.
(a) $s$-wave; (b) $d_{x^2-y^2}$-wave.
The temperatures are $T/T_{{\rm c}}=0.9$, $0.7$, $0.5$,
 and $0.3$ from top to bottom in the high-field region.}
\label{fig:MB-tau10000}
\end{center}
\end{figure}

Equation (\ref{Eilens2}) for the above model has been solved 
with the procedure of Sec.\ IID over 
$0.05H_{c2}\!\leq\! B \!\leq\! H_{c2}$.
The value $N_{{\rm cut}}$ is chosen as $30$ ($16$) for the 
$s$-wave ($d_{x^{2}-y^{2}}$-wave) pairing.
On the other hand, we have set
$\varepsilon_{{\rm c}}\!=\! 20T_{c}$ ($50T_{c}$) at $T\!=\!0.9T_{c}$ ($0.3T_{c}$).
These values have been enough to get the convergence.
Using $\Delta$, $f$, and $\tilde{A}$ thus obtained, 
we have calculated the magnetization by supercurrent, the entropy, 
and the magnetization
by Pauli Paramagnetism by Eqs.\ (\ref{eq:magnetization}), (\ref{eq:entropy}), and
(\ref{eq:spin susceptibility}), respectively.

\begin{figure}[tbp]
\begin{center}
\includegraphics[width=0.75\linewidth]{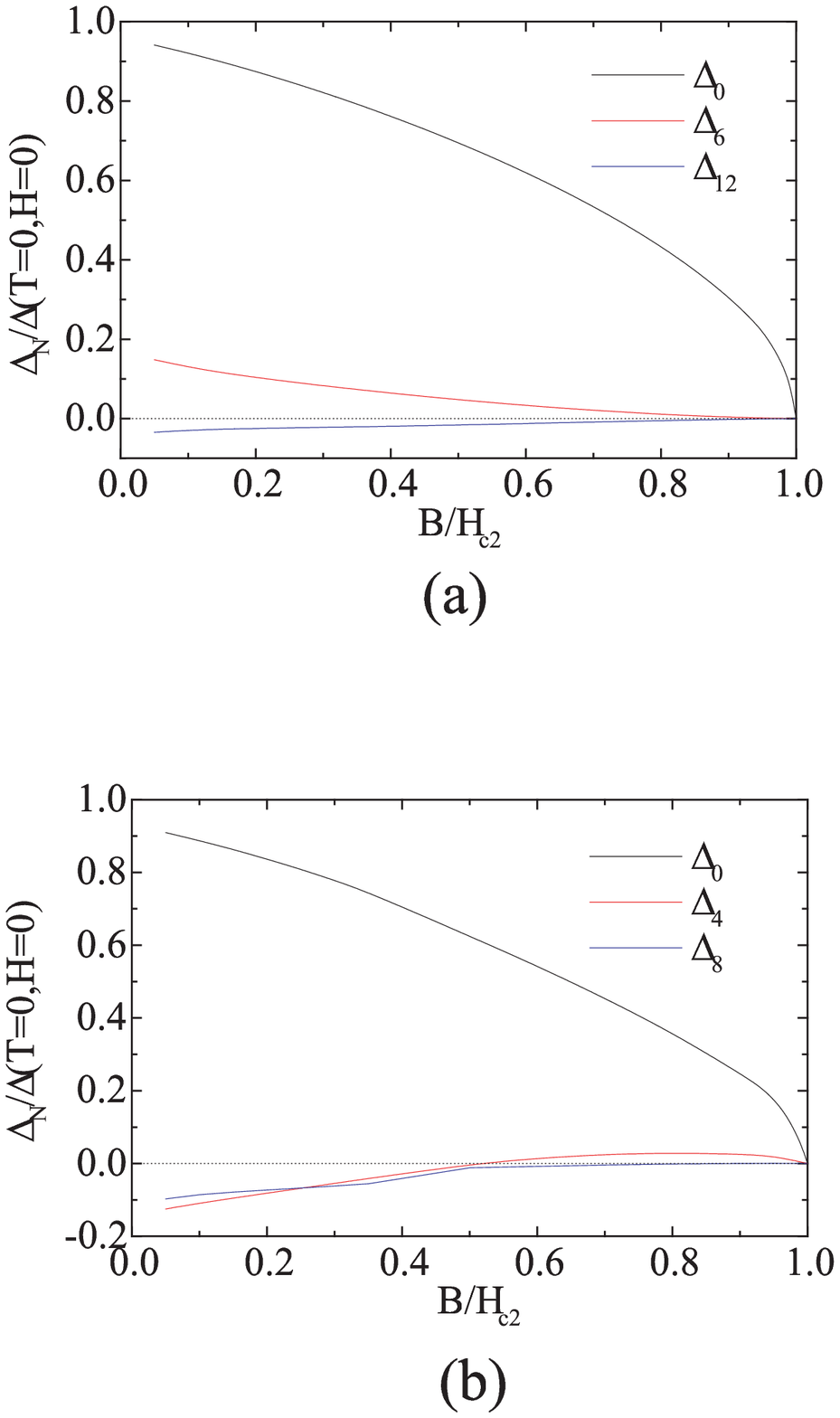}
\caption{The expansion coefficients $\Delta_{N}$ 
in Eq.\ (\ref{DExpand}) as a function of $B/H_{c2}$
in the clean limit at $T\!=\! 0.3 T_{c}$:
(a) $s$-wave pairing where the curves are
$\Delta_{0}$, $\Delta_{6}$, and $\Delta_{12}$ from top to bottom; 
(b) $d_{x^2-y^2}$-wave pairing where the curves are
$\Delta_{0}$, $\Delta_{4}$, and $\Delta_{8}$ from top to bottom
near $H_{c2}$.}
\label{fig:DeltaB}
\end{center}
\end{figure}

Figure \ref{fig:MB-tau10000}
presents magnetic-field dependence of the magnetization by supercurrent
for the $s$- and $d_{x^2-y^2}$-wave pairings in the clean limit 
at several temperatures.
In both cases, 
the initial slope at $B\!=\! H_{c2}$ gradually decreases as the temperature 
is lowered, implying a monotonic increase of 
the Maki parameter\cite{Maki64} $\kappa_{2}(T)$
as $T\!\rightarrow\! 0$.
This feature of $\kappa_{2}(T)$ has also been 
predicted in the case of the
three-dimensional spherical Fermi surface
with the $s$-wave pairing.\cite{Maki64,MT65,Eilenberger67,Kita03}
Unlike the three dimensional case,\cite{MT65,Eilenberger67,Kita03} 
however, the slope in these two-dimensional cases
remains finite and does not approach $0$
even in the clean limit of $T\!\rightarrow\!0$, 
in agreement with a previous calculation of $\kappa_{2}$.\cite{Kita03}
The curves at low temperatures become more and more
concave upward, thereby compensating the initial reduction of the magnetization.
The temperature variation is slightly 
larger for the $d_{x^2-y^2}$-wave pairing 
than the $s$-wave pairing.

\begin{figure}[tbp]
\begin{center}
\includegraphics[width=0.75\linewidth]{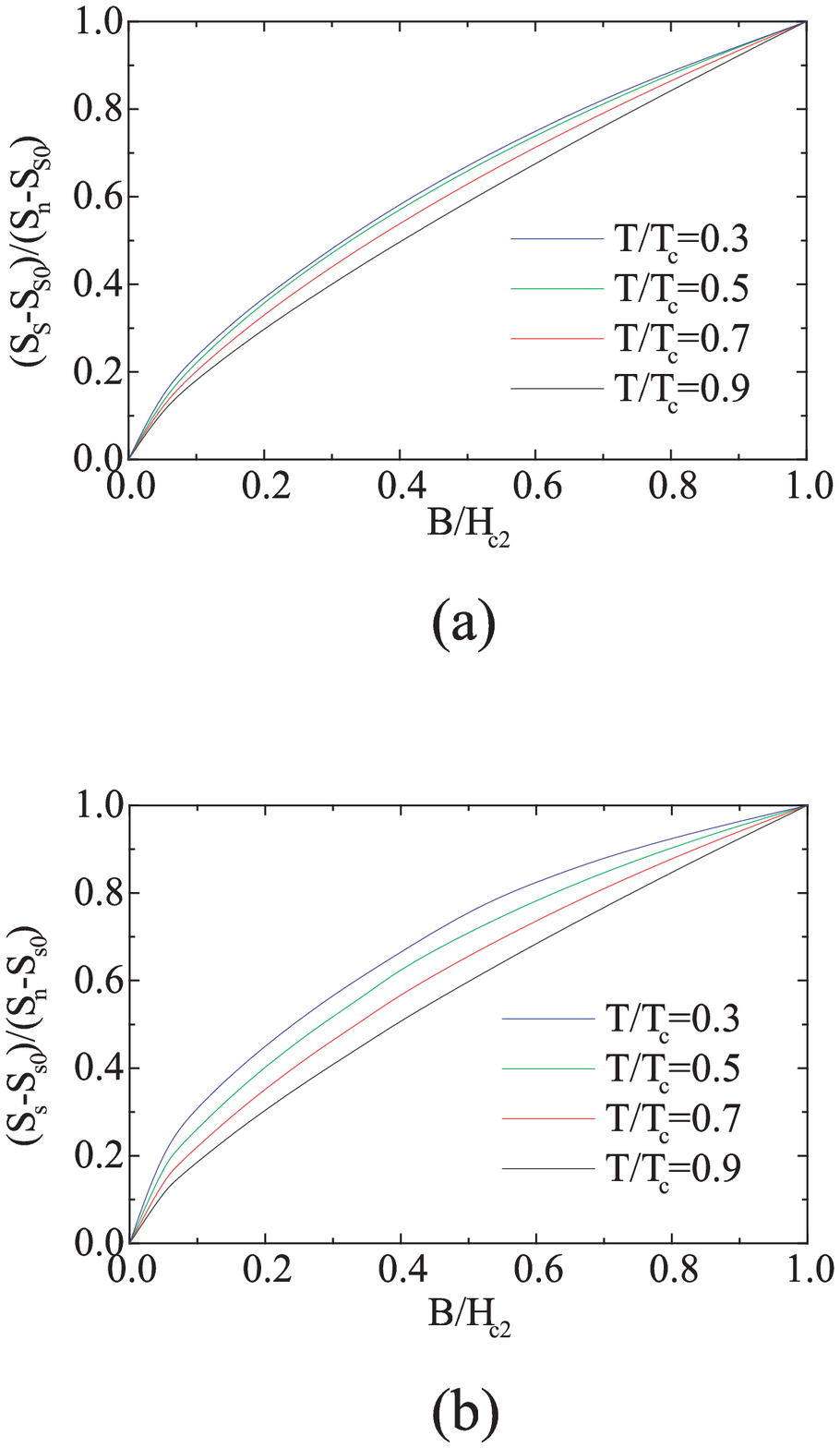}
\caption{The entropy $S_{\rm s}$ 
as a function of $B/H_{c2}$.
(a) $s$-wave; (b) $d_{x^2-y^2}$-wave.
The temperatures are $T/T_{{\rm c}}=0.9$, $0.7$, $0.5$, and $0.3$ from bottom to top.
They are normalized to vary from $1$ at
$ B\! =\! H_{c2}$ to $0$ at $B\!=\! 0$.}
\label{fig:SB-tau10000}
\end{center}
\end{figure}
\begin{table}[b]
\caption{The exponent $\alpha_{{\rm S}}$
of Eq.\ (\ref{exponentS}) 
for the $s$- and $d_{x^2-y^2}$-wave pairings
in the clean limit calculated by the best fit
to the numerical data of Fig.\ 3.}
\begin{tabular}{c|cccc}
\hline
 $T/T_{{\rm c}}$  & 0.9 & 0.7 & 0.5 & 0.3  \\
\hline
 $\alpha_{S}$ ($s$-wave)
& 0.73 & 0.69 & 0.66 & 0.63  \\
\hline
 $\alpha_{S}$ ($d$-wave)
& 0.72 & 0.66 & 0.59 & 0.52 \\
\hline
\end{tabular}
\end{table}

Figure \ref{fig:DeltaB}
shows the expansion coefficients $\Delta_{N}$
in Eq.\ (\ref{DExpand}) as a function of 
$B/H_{c2}$ for the $s$- and 
$d_{x^2-y^2}$-wave pairings in the clean limit 
at $T\!=\! 0.3T_{c}$.
Compared with the case near $T_c$,\cite{Kita98}
the mixing of higher Landau levels develops from
higher fields. 
However, the contribution is still $\sim\! 0.1\Delta_{0}$
even around $B\!=\!0.1 H_{c2}$. 
This fact implies that the Pesch approximation\cite{Pesch75}
is excellent down to $H\!\sim\!0.1 H_{c2}$ for the two-dimensional
cases with the isotropic Fermi surface.
This may not be the case for systems with complicated Fermi surfaces,
however, as suggested by the fact that 
there is already an amount of higher-Landau-level contributions
at $H_{c2}$ in those cases.\cite{RS90,Kita04,Arai04}
The Pesch approximation
may be improved to some degree by the procedure of
Graser {\em et al}.\ \cite{Graser04} 
to change $c_{1}$ and $c_{2}$ in Eq.\ (\ref{aa*})
at every temperature and magnetic field 
so that the free-energy is smallest.

\begin{figure}[tbp]
\begin{center}
\includegraphics[width=0.75\linewidth]{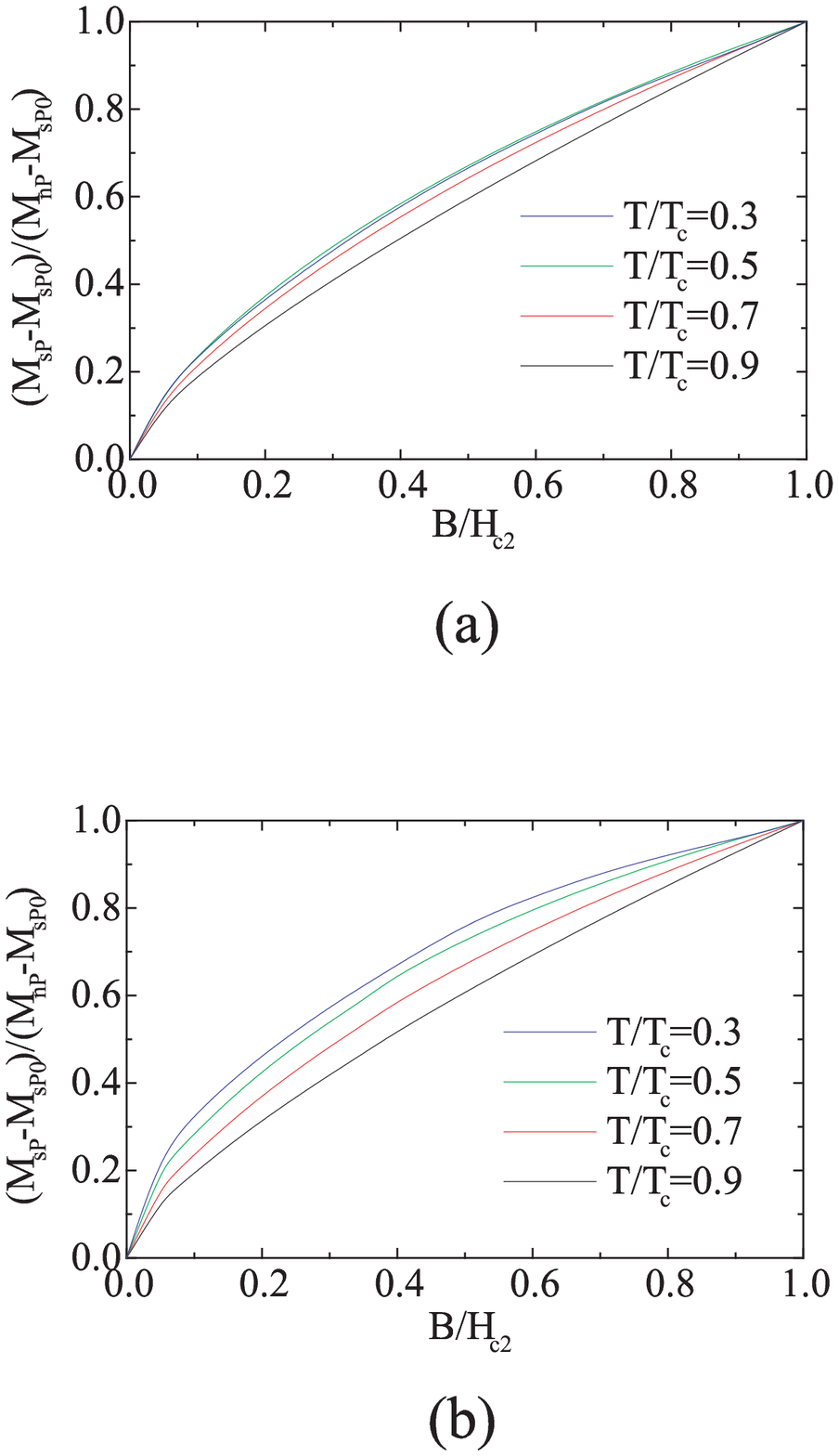}
\caption{The magnetization $M_{\rm sP}$ by Pauli paramagnetism
as a function of $B/H_{c2}$.
(a) $s$-wave; (b) $d_{x^2-y^2}$-wave.
The temperatures are $T/T_{{\rm c}}=0.9$, $0.7$, $0.5$ and $0.3$ from bottom to top.
They are normalized to vary from $1$ at
$B\! =\! H_{c2}$ to $0$ at $B\!=\! 0$.}
\label{fig:SpinB-tau10000}
\end{center}
\end{figure}
\begin{table}[b]
\caption{The exponent $\alpha_{\chi}$ 
of Eq.\ (\ref{exponentM})  
for the $s$- and $d_{x^2-y^2}$-wave pairings
in the clean limit calculated by the best fit
to the numerical data of Fig.\ 4.}
\begin{tabular}{c|cccc}
\hline
 $T/T_{{\rm c}}$  & 0.9 & 0.7 & 0.5 & 0.3 \\
\hline
 $\alpha_{\chi}$ ($s$-wave) 
& 0.71 & 0.67 & 0.63 & 0.63  \\
\hline
 $\alpha_{\chi}$ ($d$-wave)
& 0.70 & 0.63 & 0.56  & 0.49 \\
\hline
\end{tabular}
\end{table}

Figures \ref{fig:SB-tau10000} and \ref{fig:SpinB-tau10000}
plot the entropy $S_{\rm s}$ and 
the magnetization $M_{\rm sP}$
due to Pauli paramagnetism, respectively,
as a function of $B/H_{c2}$ for the $s$- and $d_{x^{2}-y^{2}}$-wave pairings
in the clean limit at several temperatures.
To see the field dependence clearly,
the entropy is normalized by using $S_{{\rm s}0}\!\equiv\!S_{\rm s}(B\!=\! 0)$
and $S_{{\rm n}}\!\equiv\!S_{\rm s}(B\!=\! H_{c2})$
as $(S_{{\rm s}}\!-\!S_{{\rm s}0})/(S_{{\rm n}}\!-\!S_{{\rm s}0})$;
it varies from $1$ to $0$ for $H_{c2}\!\geq\! B\!\geq\! 0$.
The same normalization is adopted for $M_{\rm sP}$.
All curves deviate upwards from the linear behavior $\propto\! B/H_{c2}$
and
become more and more convex upward as 
$T\!\rightarrow\! 0$.
This tendency is more conspicuous for the
$d_{x^{2}-y^{2}}$-wave pairing
due to the residual low-energy density of states.
To see the behavior quantitatively, 
we have fitted our numerical data by the formulas:
\begin{subequations}
\label{exponent}
\begin{equation}
\frac{S_{{\rm s}}\!-\!S_{{\rm s}0}}{S_{{\rm n}}\!-\!S_{{\rm s}0}}=
\left(\!\frac{B}{H_{{\rm c2}}}\!\right)^{\!\!\alpha_{\rm S}} \, ,
\label{exponentS}
\end{equation}
\begin{equation}
\frac{M_{{\rm sP}}\!-\!M_{{\rm sP}0}}{M_{{\rm n}}\!-\!M_{{\rm sP}0}}=
\left(\!\frac{B}{H_{{\rm c2}}}\!\right)^{\!\!\alpha_{\rm \chi}} \, .
\label{exponentM}
\end{equation}
\end{subequations}
To confirm the numerical results, we first estimated 
$\alpha_{S}$ and  $\alpha_{\chi}$ for the $s$-wave pairing 
by using only the data of $0.85H_{c2}\!\leq\! B\!\leq\! 0.95H_{c2}$.
Although not presented here, the procedure excellently reproduced the values of 
a previous calculation near $H_{c2}$,\cite{Kita04c}
as they should.

Table I shows the exponent
$\alpha_{S}$ obtained by the best fit to the data of 
$0.05H_{c2}\!\leq\! B\!\leq\! 0.95H_{c2}$ in Fig.\ 3.
The value of the $d_{x^2-y^2}$ pairing at $T/T_{c}\!=\!0.9$ is
almost the same as the corresponding $s$-wave result.
As the temperature is decreased, however,
the $d_{x^2-y^2}$-wave exponent decreases more rapidly
so that the curve in Fig.\ 3(b) becomes more convex upward;
this is due to the residual low-energy density of states
of the $d_{x^2-y^2}$-wave pairing.
Table II presents another exponent
$\alpha_{\chi}$ obtained by the best fit to the data of 
$0.05H_{c2}\!\leq\! B\!\leq\! 0.95H_{c2}$ in Fig.\ 4.
Each value is fairly close to the corresponding one
for $\alpha_{S}$, as may be expected from the fact that
both quantities probe the zero-energy density of states.
In this context,
Ichioka {\em et al.}\cite{Ichioka99a} calculated 
the field dependence of the zero-energy density
of states $N(0)$ at $T/T_{c}\!=\!0.5$ to find 
$N(0)\!\propto\! (B/H_{c2})^{0.67}$ and $N(0)\!\propto\! (B/H_{c2})^{0.41}$
for the $s$- and $d_{x^{2}-y^{2}}$-wave pairings, respectively.
Our estimates for $\alpha_{S}(T\!\rightarrow\!0)$ and 
$\alpha_{\chi}(T\!\rightarrow\!0)$ are somewhat smaller (larger)
for the $s$-wave ($d_{x^{2}-y^{2}}$-wave) pairing.

\begin{figure}[tbp]
\begin{center}
\includegraphics[width=0.75\linewidth]{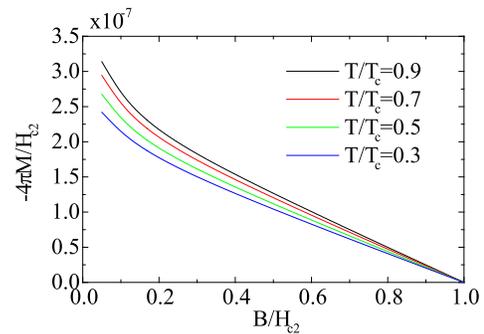}
\caption{Magnetization by supercurrent as a function of 
$B/H_{{\rm c2}}$ for the $s$-wave pairing with $\tau\!=\!\hbar/\Delta(0)$.
The temperatures are $T/T_{{\rm c}}=0.9$, $0.7$, $0.5$, 
and $0.3$ from top to bottom.}
\label{fig:MB-s-tau001}
\end{center}
\end{figure}

Experiments on the $T$-linear specific-heat coefficient $\gamma_{\rm s}(B)$ 
have been performed 
for clean V$_{3}$Si,\cite{Ramirez96}
NbSe$_{2}$,\cite{Sanchez95,Nohara99,Sonier99,Hanaguri03} and CeRu$_{2}$.\cite{Hedo98}
The quantity $\gamma_{\rm s}(B)/\gamma_{n}$ coincides for $T\!\rightarrow\! 0$ with
$(S_{s}\!-\!S_{s0})/(S_{n}\!-\!S_{s0})$ of Fig.\ 3.
Those data all show marked upward deviations
from the linear behavior $\gamma_{n}B/H_{c2}$,
indicating that it is a common feature among clean superconductors
irrespective of the energy-gap symmetry.
Sonier {\em et al.\ }\cite{Sonier99} thereby extracted the exponent 
$0.66$ for the field dependence
of $\gamma_{\rm s}(B)$ as $T\!\rightarrow\! 0$, in good agreement with
the result $0.67$ by Ichioka {\em et al.\ }\cite{Ichioka99a} for the clean 
two-dimensional $s$-wave model.
However, a more recent experiment by Hanaguri {\em et al.\ }\cite{Hanaguri03} 
reported a different exponent $0.5$.
It should also be noted that NbSe$_{2}$ has three kinds of Fermi surfaces
and one of them is quite different in structure from the cylinder.
There also exists a recent experiment which indicates existence of
different superconducting energy gaps on different Fermi surfaces.\cite{Boaknin03}
Hence the agreement between the experiment by Sonier {\em et al.\ }\cite{Sonier99}
and the theory by Ichioka {\em et al.\ }\cite{Ichioka99a}
might be an artifact and should be confirmed by more detailed experiments
as well as theories incorporating both Fermi-surface and
gap structures.
In this context, it is worth noting that no detailed experiments
have been performed on the field dependence
of $\gamma_{\rm s}(B)$ even for the classic type-II superconductors
V and Nb, although early experiments\cite{Radebaugh66,Ferreira69}
suggest similar upward deviations
from the linear behavior $\gamma_{n}B/H_{c2}$.

We next focus on the $s$-wave pairing in the dirty limit.
Figure \ref{fig:MB-s-tau001} shows the magnetization 
for $\tau\!=\!0.01\hbar/\Delta(0)$ 
as a function of $B/H_{c2}$.
Compared with the clean-limit results of Fig.\ 1(a),
we observe an extended linearity down to $B/H_{c2}\!\sim\!0.2$ 
irrespective of the temperature.
The decrease of the initial slope for $T\!\rightarrow\!0$ 
is as expected from the temperature dependence of the Maki parameter 
$\kappa_{2}$.\cite{Maki64}
By scaling this change of the initial slope, all the curves
almost fall onto a single curve.
This is a marked feature in the dirty limit which is absent in
the clean-limit result of Fig.\ 1(a).

\begin{figure}[t]
\begin{center}
\includegraphics[width=0.75\linewidth]{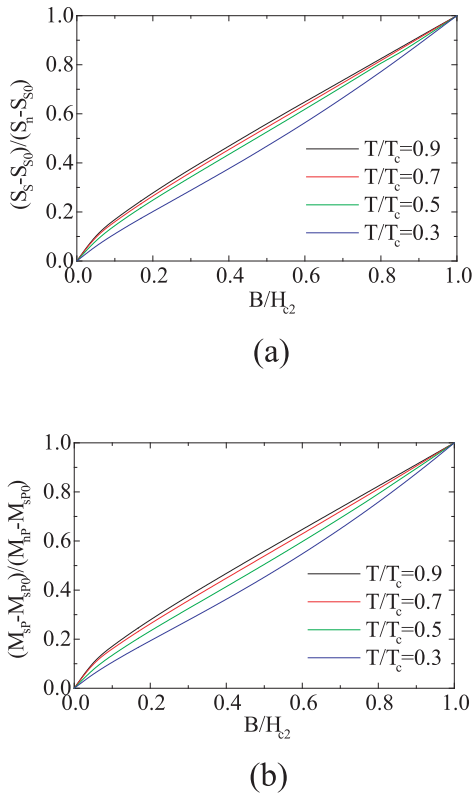}
\caption{(a) The entropy $S_{\rm S}$ and (b) the magnetization $M_{\rm sP}$ 
by Pauli paramagnetism
as a function of $B/H_{c2}$ for the $s$-wave pairing with
$\tau\!=\!0.01\hbar/\Delta(0)$.
The temperatures are $T/T_{{\rm c}}=0.9$, $0.7$, $0.5$ and $0.3$ from top to bottom. }
\label{fig:SBSpinB-s-tau001}
\end{center}
\end{figure}

\begin{table}[b]
\caption{The exponents $\alpha_{{\rm S}}$ and $\alpha_{\chi}$ 
for the $s$-wave pairing with
$\tau\!=\!0.01\hbar/\Delta(0)$ calculated by the best fit
to the numerical data of $0.5H_{{\rm c2}}\!\leq\! B\!\leq\! 0.95H_{{\rm c2}}$.}
\begin{tabular}{l|llll}
\hline
 $T/T_{{\rm c}}$  & 0.9 & 0.7 & 0.5 & 0.3 \\
\hline
 $\alpha_{S}$ 
& 0.84 & 0.88 & 0.92 & 1.10 \\
\hline
 $\alpha_{\chi}$ 
& 0.84 & 0.89 & 0.99 & 1.15 \\
\hline
\end{tabular}
\end{table}

Figure \ref{fig:SBSpinB-s-tau001} shows the field dependences of 
$S_{\rm s}$ and $M_{\rm sP}$
for $\tau\!=\!0.01\hbar/\Delta(0)$ at various temperatures.
Table III presents the corresponding
exponents $\alpha_{{\rm s}}$ and $\alpha_{{\rm \chi}}$
obtained from the data of $0.5H_{c2}\!\leq\! B\!\leq\! 0.95H_{c2}$;
unlike the clean-limit case, it has been impossible to 
fit the whole region by a single exponent, especially at intermediate temperatures,
as may be realized from Fig.\ \ref{fig:SBSpinB-s-tau001}.
Compared with Figs.\ 3(a) and 4(a), the curves are more monotonic
with the almost linear behavior $\propto\! B/H_{c2}$. 
Looking at the temperature dependence more closely, however, 
we observe a change from a convex-upward behavior at high temperatures
to a convex-downward behavior at low temperatures,
in agreement with a previous calculation near $H_{c2}$.\cite{Kita04c}
This feature also appears
in the field dependence of the
zero-energy density of states as calculated recently by Miranovi\'c
{\em et al.}\cite{Ichioka04b}
The convex-downward behavior
at $T\!=\!0.3 T_c$ may become more pronounced
at lower temperatures to be observable experimentally.

\section{Summary}

We have developed an alternative method to solve the Eilenberger equations
for the vortex-lattice state. 
The main analytic formulas are given in Sec.\ IIC together with
the numerical procedure to solve them in Sec.\ IID.
This method, which directly
extends the $H_{c2}$ equation\cite{Kita04,Arai04} to lower fields,
has a potential applicability to systems with complicated Fermi surfaces
and/or gap structures to carry out detailed calculations 
on the field dependences of thermodynamic quantities 
for various type-II superconductors.

Using it, we have calculated the
field dependences of the magnetization by supercurrent, 
the mixing of higher Landau levels in the pair potential,
the entropy, and 
the Pauli paramagnetism for the two-dimensional $s$- and $d_{x^2-y^2}$-wave pairings
in the clean and dirty limits at various temperatures.
Previous results near $H_{c2}$ for the $s$-wave pairing\cite{Kita03,Kita04c}
have been reproduced adequately
and extended to lower fields to clarify the overall field dependences.
The differences between the $s$- and $d_{x^2-y^2}$-wave pairings
are quite small at high temperatures but develop gradually 
as the temperature is lowered, reflecting the marked difference 
in the low-energy density of states between the two cases.
The field dependences of the entropy and Pauli paramagnetism
in the clean limit at low temperatures
present convex-upward behaviors for both pairings. 
In contrast, the curves of the $s$-wave pairing in the dirty limit 
are more monotonic and fairly close to the linear behavior,
but also acquire downward curvature at low temperatures.
As for the magnetization by supercurrent, there is a wide region
of linear field dependence from $H_{c2}$ both at high temperatures 
and in the dirty limit. The region shrinks in the clean limit 
as the temperature is lowered, and the curve acquires pronounced
upward curvature.
It is also found that the mixing of higher Landau levels in the pair potential
is small for $B\!\agt\! 0.1H_{c2}$ but develops rapidly as the field is
further decreased.

\begin{acknowledgments}
This work is supported by a Grant-in-Aid for Scientific Research from the Ministry
of Education, Culture, Sports, Science, and Technology of Japan, and also from
the 21st century COE program ``Topological Science and Technology,'' 
Hokkaido University.
\end{acknowledgments}


\end{document}